\documentclass[11pt]{article}

\usepackage[margin=1in]{geometry}
\usepackage[utf8]{inputenc}
\usepackage{microtype}
\usepackage{amsmath,amssymb,amsthm}
\usepackage{graphicx}
\usepackage{booktabs}
\usepackage{tabularx}
\usepackage{multirow}
\usepackage{array}
\usepackage{xcolor}
\usepackage{tikz}
\usetikzlibrary{arrows.meta,positioning,fit,calc,backgrounds}
\usepackage{algorithm}
\usepackage{algpseudocode}
\usepackage[numbers,sort&compress]{natbib}
\usepackage[hidelinks]{hyperref}
\usepackage{caption}
\usepackage{float}
\theoremstyle{definition}
\newtheorem{definition}{Definition}

\newcommand{\sys}{\textsc{Frontier Autolab}}
\newcolumntype{Y}{>{\raggedright\arraybackslash}X}

\title{Frontier Autolab: Organizational Memory, Adversarial Dissent and Temporal Leakage in Multi-Agent LLM Firms Across Fifty Years of Technological Change}
\author{Bravish Ghosh\\
\small Independent researcher}
\date{September 2026}

\begin{document}
\maketitle

\begin{abstract}
Multi-agent LLM systems are increasingly structured like organizations, with roles, critics and shared memory, yet they are evaluated on tasks that last minutes. We ask how such an organization behaves when the ground it stands on keeps moving. \sys{} is a long-horizon testbed in which one simulated firm, voiced by sixteen role personas and a dedicated Red Team, must re-found itself in nine technology eras from 1990 to 2040. Each era is temporally gated: the firm decides from a dated briefing, a historian-judge then reveals what happened and scores the decision on a five-dimension rubric, and lessons enter a persistent Playbook. Six eras are scored against history, one against the live market and two are open forecasts. Across four trajectories (36 era decisions, 180 subscores) we find a consistent \emph{foresight--commitment gap}: in all 24 historically scored eras the judge rated the firm's recognition of the coming shift above its choice of where to build (mean gap 1.9 points on a 10-point scale), because boards chose the layer their existing assets could reach. Organizational design shaped long-run character. A Red Team armed with numeric kill gates produced fifty years of gated pilots and no product, and the rubric rated this firm highest; firms whose memory stored market-structure lessons pivoted every era, while a firm whose memory stored only validation procedure kept one method throughout. We also show why such results are hard to trust. Scores rise across eras in every run while the judge's own hindsight subscore falls (within-run $r=-0.58$), so apparent learning is confounded with recall of history, and we trace further distortions to self-judging, briefing selection and score aggregation. We release all records and an API harness, and specify fictional and post-cutoff eras that would turn the testbed into a benchmark.
\end{abstract}

\section{Introduction}

A firm that was right about one decade of technology is rarely right about the next. Management research has spent decades on why: organizations search near what they already know \citep{cohen1990absorptive}, exploitation crowds out exploration \citep{march1991exploration}, and incumbents rationally serve the market they have instead of the one that is forming \citep{christensen1997innovator}. LLM agents are now being assembled into organizations of their own, with executives, engineers, reviewers and shared memory \citep{park2023generative,qian2024chatdev,hong2024metagpt,li2023camel,chen2024agentverse,guo2024survey}. These systems are evaluated on the time scale of a task: a repository, a debate, a game. Whether they inherit, avoid or amplify the long-run failures of human firms is unknown, because no evaluation runs long enough to see them.

We build such an evaluation. \sys{} places one simulated company in nine successive technology eras, from office networking in 1990 to delegated AI agents in 2040, with a single mandate: become the most important company each era could produce, even if that means ending the current business. In every era three departments write strategy memos, a Red Team attacks them, a board decides what the company becomes, and a judge scores the decision. For the six eras between 1990 and 2020 the judge scores the firm against what actually happened; the live era is audited against the present market; the last two eras are forecasts. Lessons accumulate in a Playbook that the firm must cite in every later era.

History is an appealing answer key, and a treacherous one. The model voicing the firm already knows which companies won, and so does the model judging it \citep{sainz2023trouble,golchin2024timetravel,cheng2024dated,glasserman2023lookahead,paleka2026pitfalls}. We therefore treat measurement as a first-class object of study rather than an afterthought.

\paragraph{Findings.} Across four trajectories run under three orchestration schemes:

\begin{enumerate}
  \item \textbf{Foresight--commitment gap.} In all 24 historically scored eras, the judge rated \emph{frontier accuracy} above \emph{layer choice}. Memos named the coming shift (hypertext in 1990, click-based ranking in 1996, neural-network scale in 2020), but boards built the layer their assets could reach (Section~\ref{sec:gap}).
  \item \textbf{Critics can freeze a firm.} A Red Team that issued numeric kill gates in every era produced a company that ran gated manual pilots for fifty years and never shipped a product. Because the rubric rewarded discipline and penalized hindsight, this firm scored highest (Section~\ref{sec:caution}).
  \item \textbf{Memory content sets character.} Firms whose Playbook held lessons about technology and market structure changed identity every era; a firm whose Playbook held only validation procedure changed identity three times in nine eras (Section~\ref{sec:memory}).
  \item \textbf{Apparent learning is confounded with recall.} Scores rise across eras in every run while the judge's hindsight subscore falls in the same eras. Self-judging, briefing selection and inconsistent score aggregation distort the picture further (Section~\ref{sec:validity}).
\end{enumerate}

\paragraph{Contributions.} (i) A formulation of long-horizon organizational evaluation as a sequence of temporally gated strategic decisions with explicit memory and dissent (Section~\ref{sec:formulation}). (ii) \sys{}, an open testbed implementing it: role charter, per-era protocol, rubrics and an API harness with ablations (Section~\ref{sec:system}). (iii) Four complete trajectories with every briefing, memo, board decision and reveal that was produced, one disclosed operator intervention and all redactions. (iv) Behavioural findings on LLM organizations, connected to the organizational-learning literature. (v) An audit of historically scored agent evaluation and a benchmark design with fictional and post-cutoff eras, harness-computed scores and human raters.

\section{Related work}

\paragraph{LLM agent organizations.} Role-playing frameworks assign LLM instances roles and protocols to write software \citep{qian2024chatdev,hong2024metagpt}, cooperate on tasks \citep{li2023camel,chen2024agentverse} or simulate social life \citep{park2023generative}; \citet{guo2024survey} survey the area. Multi-agent debate improves factuality and diversity of reasoning \citep{du2024debate,liang2024divergent}, which motivates red-team roles. Agents that accumulate verbal lessons or skills across episodes \citep{shinn2023reflexion,wang2023voyager} are the closest analogue to our Playbook. These works measure task success over short horizons; we measure a chain of strategic decisions over five simulated decades. Work on the length of tasks agents can complete \citep{kwa2025timehorizon} measures horizon in working hours; our horizon is organizational.

\paragraph{LLMs as simulated people and firms.} LLMs have been used as simulated economic agents \citep{horton2023homo} and survey respondents \citep{argyle2023outofone}. As in that work, we treat the firm's outputs as model behaviour, not evidence about real companies.

\paragraph{Forecasting and temporal leakage.} Forecasting benchmarks score probabilistic answers to resolvable questions \citep{zou2022autocast,halawi2024forecasting,karger2024forecastbench}, in the tradition of \citet{tetlock2005expert,tetlock2015superforecasting}. Strategic decisions do not resolve cleanly, so we rely on a judge. Retrospective evaluation of LLMs suffers from contamination \citep{sainz2023trouble,golchin2024timetravel}, uncertain effective cutoffs \citep{cheng2024dated} and look-ahead bias \citep{glasserman2023lookahead}; \citet{paleka2026pitfalls} catalogue these pitfalls for LLM forecasters. We add two channels specific to organizational simulation: the choice of what a dated briefing mentions, and a judge that labels foresight after the fact.

\paragraph{LLM judges.} LLM judges track human preferences on many tasks but show position and verbosity biases \citep{zheng2023judging,wang2023notfair} and prefer their own generations \citep{panickssery2024selfpref}. In each of our runs the judge is the same model as the players.

\paragraph{Organizational learning.} Our protocol operationalizes organizational learning as the encoding of experience into routines \citep{levitt1988organizational}, exploration versus exploitation \citep{march1991exploration}, absorptive capacity \citep{cohen1990absorptive}, dynamic capabilities \citep{teece1997dynamic} and disruption \citep{christensen1997innovator}. The Red Team follows devil's-advocacy research \citep{schwenk1990devils} and is a guard against groupthink \citep{janis1972groupthink}. Hindsight bias in human judgment \citep{fischhoff1975hindsight,roese2012hindsight} is the human counterpart of the leakage we measure.

\section{Problem formulation}
\label{sec:formulation}

\paragraph{Eras and information sets.} A trajectory is a sequence of eras $k=1,\dots,K$ with start dates $t_1<\dots<t_K$. Let $\mathcal{W}$ denote the world's record of events. The \emph{admissible information} for era $k$ is $\mathcal{I}_k=\{w\in\mathcal{W}:\text{date}(w)<t_k\}$, and the era's \emph{outcome window} is $\mathcal{O}_k=\{w:t_k\le\text{date}(w)<t_{k+1}\}$. A briefing $B_k$ is a selection from $\mathcal{I}_k$ made by a briefing author.

\paragraph{The organization.} An organization is a set of roles $\mathcal{R}$, each a persona with a lens, together with a memory $P_k$ (the Playbook) and a company state $S_k$ (assets, capital, identity). A decision procedure $\Pi$ maps $(B_k,P_k,S_k)$ to a decision
\begin{equation}
D_k=(\text{identity},\ \text{thesis},\ \ell_k,\ \text{kills},\ \text{wedge},\ \text{kill criteria},\ \text{dissent log}),
\end{equation}
where $\ell_k$ is the layer of the technology stack the firm chooses to own. The procedure is itself structured: departments propose, a critic attacks, an executive decides.

\paragraph{Judging and memory.} A judge $J$ observes $D_k$ and, in historically scored eras, $\mathcal{O}_k$. It returns subscores $\mathbf{s}_k\in\{0,\dots,10\}^5$, an outcome that updates $S_{k+1}$, and lessons $L_k$ with $P_{k+1}=P_k\cup L_k$. The judge also authors $B_{k+1}$.

\paragraph{Leakage.} The procedure $\Pi$ is implemented by a model with parametric knowledge of $\mathcal{W}$ up to its training cutoff. \emph{Hindsight leakage} occurs when $D_k$ depends on $\mathcal{O}_k$ through that knowledge rather than through $B_k$. Leakage can also enter through $B_k$ itself: every fact in $B_k$ may lie in $\mathcal{I}_k$ while its \emph{selection} depends on $\mathcal{O}_k$. We call this \emph{briefing-selection leakage}.

\paragraph{Quantities of interest.} Beyond the era total, we define two behavioural measures.

\begin{definition}[Foresight--commitment gap]
For a historically scored era, $G_k=s_k^{\text{frontier}}-s_k^{\text{layer}}$, the judge's rating of how well the organization identified the coming capability shift minus its rating of the layer the organization chose to own. $G_k>0$ means the organization saw more than it acted on.
\end{definition}

\begin{definition}[Identity pivot]
A pivot occurs in era $k>1$ if the board adopts a company identity different from era $k-1$. The pivot rate of a trajectory is the fraction of its $K-1$ transitions that are pivots.
\end{definition}

\section{The \sys{} testbed}
\label{sec:system}

\subsection{Eras and mandate}

\sys{} instantiates the formulation with $K=9$ (Table~\ref{tab:eras}). Training eras E1--E6 are scored against history over roughly six-year windows. The live era E7 (September 2026) is audited against the current market with web search. Forecast eras E8 (2032) and E9 (2040) have no answer key; the judge audits them for internal consistency and plausibility. The firm's only mandate is \emph{reinvention toward the frontier}.

\begin{table}[t]
\centering
\small
\caption{Eras. The E6 outcome window extends to the live era and is longer than the others.}
\label{tab:eras}
\begin{tabular}{llll}
\toprule
Era & Start $t_k$ & Mode & Scored against \\
\midrule
E1 & Jan 1990 & training & history, 1990--1995 \\
E2 & Jan 1996 & training & history, 1996--2001 \\
E3 & Jan 2002 & training & history, 2002--2007 \\
E4 & Jan 2008 & training & history, 2008--2013 \\
E5 & Jan 2014 & training & history, 2014--2019 \\
E6 & Jan 2020 & training & history, 2020--Aug 2026 \\
E7 & Sep 2026 & live & current market, checked with web search \\
E8 & Jan 2032 & forecast & projected world (no ground truth) \\
E9 & Jan 2040 & forecast & projected world (no ground truth) \\
\bottomrule
\end{tabular}
\end{table}

\subsection{Roles}

The charter defines sixteen personas in five groups and two external judges (Table~\ref{tab:roles}). A persona is a role with a lens that a model call adopts. How personas map to model contexts differs by run (Section~\ref{sec:setup}).

\begin{table}[t]
\centering
\small
\caption{Role charter.}
\label{tab:roles}
\begin{tabularx}{\linewidth}{lcY}
\toprule
Group & Personas & Lens \\
\midrule
Executive & 4 & CEO (final call), CTO (feasibility), Chief Scientist (which curves bend), CSO (where value pools) \\
Frontier Research & 4 & Emerging capabilities, historical analogies, scenarios, what cannot yet be measured \\
Product \& Engineering & 4 & What a small team can ship, the wedge, unavoidable infrastructure, ecosystems \\
Market \& Capital & 3 & Who pays first, financing climate, cost curves and margins \\
Governance & 1 & Red Team: attacks every proposal, hunts hype and hindsight \\
External judges & 2 & The Record (training eras), The Auditor (live and forecast eras) \\
\bottomrule
\end{tabularx}
\end{table}

\subsection{Per-era protocol}

\begin{figure}[t]
\centering
\begin{tikzpicture}[
  node distance=6mm and 7mm,
  box/.style={draw, rounded corners=3pt, align=center, font=\small, minimum height=10mm, inner sep=4pt, fill=white},
  acc/.style={box, draw=blue!60!black, fill=blue!6, line width=0.9pt},
  jud/.style={box, draw=orange!70!black, fill=orange!8},
  arr/.style={-{Stealth[length=2.2mm]}, thick, draw=black!65}
]
\node[jud, text width=24mm] (brief) {World briefing $B_k$\\\scriptsize facts dated $<t_k$};
\node[box, text width=25mm, right=10mm of brief, yshift=12mm] (fr) {Frontier Research};
\node[box, text width=25mm, right=10mm of brief] (pe) {Product \& Eng.};
\node[box, text width=25mm, right=10mm of brief, yshift=-12mm] (mc) {Market \& Capital};
\node[acc, text width=24mm, right=10mm of pe] (board) {Board $\to D_k$\\\scriptsize Red Team attacks,\\\scriptsize CEO decides};
\node[jud, text width=24mm, right=8mm of board] (rev) {Reveal \& score\\\scriptsize $\mathbf{s}_k$ vs.\ $\mathcal{O}_k$};
\node[box, text width=62mm, below=15mm of board, xshift=-6mm] (pb) {Playbook $P_{k+1}$ + company state $S_{k+1}$};
\node[font=\scriptsize, above=0.5mm of fr] {3 memos in parallel};
\draw[arr] (brief.east) -- ++(4mm,0) |- (fr.west);
\draw[arr] (brief.east) -- (pe.west);
\draw[arr] (brief.east) -- ++(4mm,0) |- (mc.west);
\draw[arr] (fr.east) -- ++(4mm,0) |- (board.west);
\draw[arr] (pe.east) -- (board.west);
\draw[arr] (mc.east) -- ++(4mm,0) |- (board.west);
\draw[arr] (board.east) -- (rev.west);
\draw[arr] (rev.south) |- (pb.east);
\draw[arr] (pb.west) -| node[pos=0.25, above, font=\scriptsize] {next era} (brief.south);
\end{tikzpicture}
\caption{One era of \sys{}. Orange boxes are written by the judge, which both scores era $k$ and selects the facts in $B_{k+1}$; this dual role is one of the leakage channels analysed in Section~\ref{sec:validity}.}
\label{fig:loop}
\end{figure}
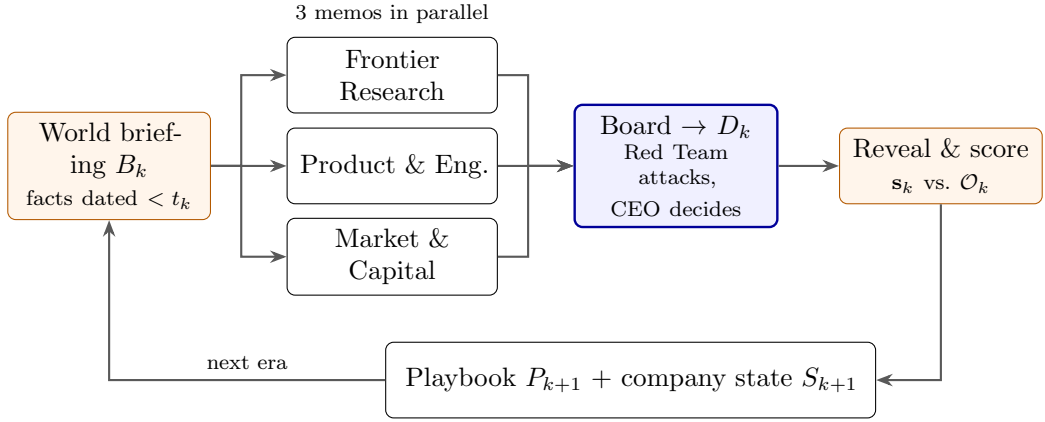

Figure~\ref{fig:loop} and Algorithm~\ref{alg:loop} give one era. Three departments write memos in parallel. Each memo proposes two theses, each stated as the chain \emph{capability $\rightarrow$ adoption $\rightarrow$ bottleneck $\rightarrow$ layer owned $\rightarrow$ proprietary data $\rightarrow$ next capability}. At the board meeting the Red Team attacks every thesis, and the CEO issues $D_k$, including what the company kills and a log of dissent. The judge then reveals and scores, writes lessons to the Playbook, updates the company state and writes the next briefing.

\begin{algorithm}[t]
\caption{One \sys{} trajectory}
\label{alg:loop}
\small
\begin{algorithmic}[1]
\Require charter $\mathcal{R}$, eras $E_1,\dots,E_K$, player model $\pi$, judge model $J$
\State $S_1 \gets$ seed company (\$1.5M, 16 personas); $P_1 \gets \emptyset$
\State $B_1 \gets J.\textsc{Briefing}(E_1)$ \Comment{selected from $\mathcal{I}_1$}
\For{$k = 1$ to $K$}
  \For{$d \in \{\text{frontier}, \text{product}, \text{market}\}$ \textbf{in parallel}}
    \State $M_d \gets \pi.\textsc{Memo}(d, B_k, P_k, S_k, D_{k-1}, R_{k-1})$ \Comment{two theses; cites $P_k$}
  \EndFor
  \State $C_k \gets \pi.\textsc{RedTeam}(\{M_d\}, P_k)$
  \State $D_k \gets \pi.\textsc{CEO}(\{M_d\}, C_k, P_k, S_k)$
  \If{$E_k$ is a training era}
    \State $(R_k, \mathbf{s}_k, L_k) \gets J.\textsc{Reveal}(D_k, \mathcal{O}_k)$
  \Else
    \State $(R_k, \mathbf{s}_k, L_k) \gets J.\textsc{Audit}(D_k)$ \Comment{web search in E7}
  \EndIf
  \State $P_{k+1} \gets P_k \cup L_k$; \quad $S_{k+1} \gets S_k \oplus \text{outcome}(R_k)$
  \State $B_{k+1} \gets J.\textsc{Briefing}(E_{k+1})$
\EndFor
\State \Return $\{D_k, R_k, \mathbf{s}_k\}_{k=1}^{K}$, $P_{K+1}$
\end{algorithmic}
\end{algorithm}

\paragraph{Rules.} \emph{Anti-hindsight}: in training eras, memos and decisions may use only $\mathcal{I}_k$; the Red Team polices this and the judge penalizes violations. \emph{Reinvention by default}: keeping the previous business must be argued for. \emph{Layer thinking}: every thesis states the full chain above. \emph{Small-team realism}: the wedge must be buildable by fewer than twenty people with era-realistic capital. \emph{Playbook primacy}: from E2 on, every memo cites a Playbook lesson or argues against one.

\paragraph{Rubrics.} Training eras are scored on frontier accuracy, timing, layer choice, reinvention courage and hindsight leakage (10 = none). Live and forecast eras are scored on playbook consistency, plausibility, non-consensus, layer choice and groundedness. The judge also reports a total $T_k\in[0,100]$. Because the protocol did not fix how $T_k$ relates to the subscores, we also report the mechanical aggregate $\tilde{T}_k=2\sum_{j=1}^{5}s_{k,j}$.

\section{Experimental setup}
\label{sec:setup}

We ran four trajectories on 28--29 September 2026 (Table~\ref{tab:runs}). They differ in orchestration, which lets us see which behaviours survive a change of scheme, but they are not a controlled sample.

\begin{table}[t]
\centering
\small
\caption{The four trajectories.}
\label{tab:runs}
\begin{tabularx}{\linewidth}{lYYY}
\toprule
Run & Orchestration & Published per-era records & Known issues \\
\midrule
001 & About 45 separate agent launches of Claude Opus 5.5, each starting from shared files only; one launch per memo, board and judgment & Briefing, 3 memos, board decision, reveal; Playbook & E6 board decision written by the operator during a service outage \\
002 & One Luna 6 context in the Codex agent voiced all roles and the judge & Synthesis, Playbook, company state, scores & Probable exposure to Run 001; export column shift \\
003 & As Run 002 & Short decision and reveal per era; synthesis & As Run 002; E6 window ends 2025 \\
004 & Luna 6 in the Codex agent, with separate persistent contexts per department and Red Team; root context wrote briefings, decisions and judgments & Briefing, 3 memos, Red Team critique, board decision, reveal & Probable exposure to Run 001; judge not independent of the decision-maker \\
\bottomrule
\end{tabularx}
\end{table}

\paragraph{Models.} Run 001 used Claude Opus 5.5 in an agent environment that could launch sub-agents with file and web tools. Runs 002--004 used Luna 6 through the OpenAI Codex agent. In every run the same model voiced all personas and the judges. Sampling settings were those of each agent environment and were not recorded. The API harness released with the testbed, which separates player and judge models and supports ablations, was not used for these runs.

\paragraph{Operator intervention.} During Run 001 the service that approves new agent launches timed out for about thirty minutes at the E6 board step. The operator wrote the E6 board decision from the E6 memos, Playbook and company state and told the E6 judge, which penalized the era for the missing Red Team. We exclude nothing from the analysis but flag E6 of Run 001 wherever it matters.

\section{Results: how LLM firms behave over fifty years}
\label{sec:results}

\subsection{Overview}

\begin{figure}[t]
\centering
\includegraphics[width=\linewidth]{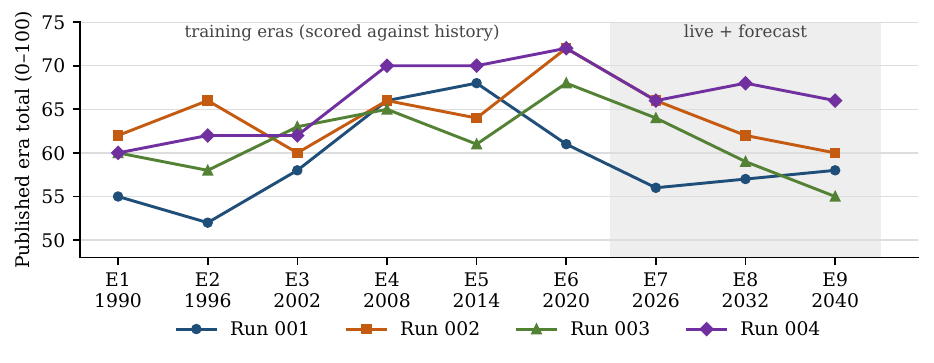}
\caption{Published era totals. Training and live/forecast eras use different rubrics and should be compared only within mode.}
\label{fig:traj}
\end{figure}

Figure~\ref{fig:traj} shows the published totals. All 36 lie between 52 and 72 (mean 62.7, SD 5.0). In every run the training-era slope is positive (+1.4 to +2.6 points per era) and the last three training eras beat the first three by 4.3 to 10.0 points (Table~\ref{tab:summary}). Section~\ref{sec:validity} asks whether this rise means anything; here we describe behaviour.

\subsection{The foresight--commitment gap}
\label{sec:gap}

\begin{figure}[t]
\centering
\includegraphics[width=\linewidth]{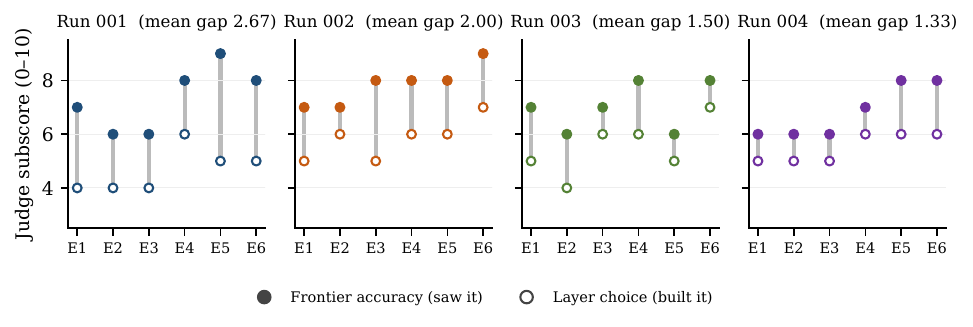}
\caption{Foresight--commitment gap. For each training era, the judge's frontier-accuracy subscore (filled) and layer-choice subscore (open). The gap is positive in 24 of 24 eras.}
\label{fig:gap}
\end{figure}

The gap $G_k$ is positive in \emph{all} 24 training eras across the four runs (mean 1.88, SD 0.80; Figure~\ref{fig:gap}). Layer choice is the lowest mean training subscore in every run (4.67, 5.83, 5.50, 5.50), while frontier accuracy is the highest (7.33, 7.83, 7.00, 6.83). If the gap were rubric noise, its sign would vary; under a sign test it would be positive in 24 of 24 eras with probability below $10^{-7}$. That test treats eras as independent, which they are not, and in each run the judge is the same model as the players, so we read it as a strong regularity of the judged behaviour rather than a population estimate.

Run 001's full records show the mechanism (Table~\ref{tab:adjacent}). In five of six eras a department memo named the shift the judge later identified as decisive, and the board chose instead a layer adjacent to the firm's existing assets: a mail gateway rather than the browser and index in 1990, a receipted EDI exchange rather than search in 1996, a moderation API rather than compute and data in 2014. The research memo often held the right idea as a funded option or a watch item, and the board killed it when it did not fit the current business; in 1996 click-based ranking was kept as research and killed in 1998. E3 is the exception in which no memo named the era's largest shifts (social networks, cloud and mobile). This is the pattern that absorptive-capacity and exploration--exploitation theories predict for human firms \citep{cohen1990absorptive,march1991exploration}: firms perceive distant opportunities but commit near existing competences.

\begin{table}[t]
\centering
\small
\caption{Run 001: what the memos named versus what the board built, as labelled by the judge.}
\label{tab:adjacent}
\begin{tabularx}{\linewidth}{lYYY}
\toprule
Era & Named in a memo & Board's choice & Where value pooled (judge) \\
\midrule
E1 1990 & Hypertext and an index across networks & Mail and address gateway & OS, routers, access providers; then browser and index \\
E2 1996 & Click-based relevance ranking (killed 1998) & Receipted Web-EDI exchange & Search and paid listings \\
E3 2002 & Contextual matching (funded option) & Cross-channel conversion ledger & Auction owners; social, cloud, mobile (not named) \\
E4 2008 & Neural-network revival (watch item) & In-app action exchange & Platform-owned app-install auctions \\
E5 2014 & Deep learning on GPUs & General moderation API & Chips, compute, labelled data \\
E6 2020 & Scale of pretrained models & Outcome-graded evaluation$^\dagger$ & Frontier labs, compute, expert data \\
\bottomrule
\end{tabularx}
\par\smallskip\raggedright\footnotesize $^\dagger$Board decision written by the operator.
\end{table}

\subsection{Critics can freeze a firm}
\label{sec:caution}

Run 004 is the one trajectory in which the Red Team ran in its own persistent context. It issued numeric kill thresholds in every era (for example, ``stop if fewer than 3 of 10 support managers confirm recurring incidents''), and every board adopted them as gates: a manual pilot, no software build until the gates pass. Over fifty simulated years the firm moved from LAN support records to online-order exceptions, payment-dispute evidence and independent workflow qualification. It treated the Web and learned models only as deferred options and never built a product.

This firm has the highest training mean (66.0), the highest hindsight subscores (mean 7.33 against 5.17--6.00), and 9/10 for playbook consistency in both forecast eras, for re-applying the same gates. Its groundedness was 8/10 and its non-consensus 5/10 in all three live and forecast eras: the judge found it reliable and unremarkable. The rubric rewards the absence of hindsight and the presence of discipline, and nothing in it asks whether the company reached where value pooled. We call the result a \emph{caution attractor}: a strong critic, combined with a scorer that penalizes bold claims more than missed opportunities, drives the organization to a stable state of inaction that is scored as prudence. Human organizations show the same drift under loss framing \citep{kahneman1993timid}, and devil's advocacy is known to improve scrutiny while slowing decisions \citep{schwenk1990devils}. The LLM firm reached this state in one era and never left it.

\subsection{What the organization remembers}
\label{sec:memory}

The Playbook is the firm's only memory beyond its company state, and its content differed sharply across runs (Table~\ref{tab:memory}). Run 001 accumulated about forty lessons, mostly about technology and market structure (``bridges expire when the standards war ends'', ``the auction owner absorbs the measurement'', ``a public curve pools value at its scarce inputs''), and revised them when history disagreed: the E3 judge softened ``suspect comfort'' to ``judge on market pull'' after comfortable options it had rejected became large businesses. Run 004's fifteen lessons are all about validation procedure (measure a baseline, find the payer, set kill thresholds before testing, require a second paying buyer); none names a technology or a market structure.

\begin{table}[t]
\centering
\small
\caption{Memory content and identity change. Pivot rate is the share of the eight era transitions in which the company changed identity (Appendix~\ref{app:identity}).}
\label{tab:memory}
\begin{tabularx}{\linewidth}{lcYc}
\toprule
Run & Playbook size & Dominant lesson type & Pivot rate \\
\midrule
001 & $\approx$40, with revisions & Technology and market structure & 8/8 \\
002 & 9 (one per era) & Market structure and buyers & 8/8 \\
003 & 9 (one per era) & Buyers, regulated workflows and evidence & 8/8 \\
004 & 15 & Validation procedure only & 3/8 \\
\bottomrule
\end{tabularx}
\end{table}

Behaviour followed memory. The three firms whose lessons described markets re-founded themselves at every transition; the firm whose lessons described procedure kept the Switchyard name for six eras and changed its business only three times, each time as an extension of the same method. Organizational-learning theory describes firms as encoding experience into routines that then govern action \citep{levitt1988organizational}. The LLM firms made that encoding explicit: what the judge wrote into memory in one era became the frame the departments argued within in the next. Memory in agent systems is usually evaluated as a performance aid \citep{shinn2023reflexion,wang2023voyager}; here its \emph{content} set the organization's character.

Run 001 also logged dissent at every board meeting, and its final synthesis identifies vindicated dissents in every era from E1 to E8, most often the voice asking who holds the money or bears the loss. Wrong dissents were not counted, so this is not yet an accuracy rate; we include it in the benchmark metrics below.

\subsection{Convergent futures}

By 2040 all four firms had arrived at related theses about accountability for delegated AI work (Table~\ref{tab:converge}): bonding, recourse, warranty or qualification for actions taken by agents without a human signature. The later runs were more sceptical than Run 001 that software evidence alone creates an insurable market; Runs 002 and 003 required a licensed risk-bearing partner and calibrated loss data first. The convergence spans two models, but Runs 002--004 share one model and were probably exposed to Run 001, so it is best read as a model prior about where value pools once agents act autonomously, not as independent evidence about the future.

\begin{table}[t]
\centering
\small
\caption{Live and forecast theses.}
\label{tab:converge}
\begin{tabularx}{\linewidth}{lYYY}
\toprule
Run & E7 (Sep 2026) & E8 (2032) & E9 (2040) \\
\midrule
001 & Long-horizon training environments from real enterprise work & Clearing house for delegated agent work & Bonding house for agents acting without a human signature \\
002 & Rights-bearing workflow traces and reliability evidence & Acceptance records for delegated work & Bounded recourse for delegated actions \\
003 & Acceptance tests for one AI claims workflow & Scoped, revocable authority for delegated tasks & Capped recourse for one delegated transaction class \\
004 & Independent qualification of one dispute workflow & Paid manual acceptance testing & Decision-linked evaluation \\
\bottomrule
\end{tabularx}
\end{table}

\section{Can the scores be trusted?}
\label{sec:validity}

Everything above is seen through the judge. This section examines the judge.

\subsection{Learning or remembering?}
\label{sec:leakage}

\begin{figure}[t]
\centering
\includegraphics[width=\linewidth]{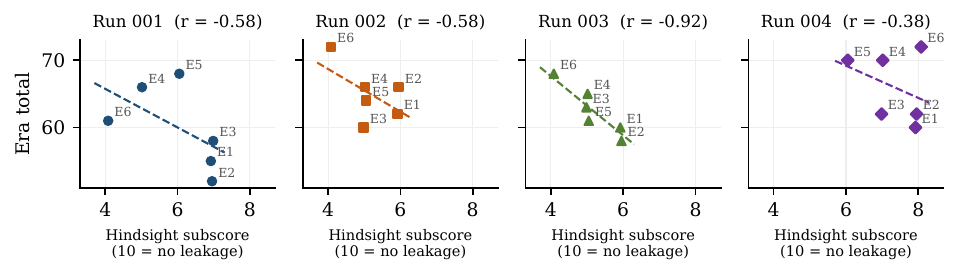}
\caption{Training-era total against the judge's hindsight subscore. Dashed lines are least-squares fits; labels mark eras.}
\label{fig:leak}
\end{figure}

If the Playbook made the firm better, scores should rise over eras, and they do. But the judge's hindsight subscore falls in the same eras. Every run shows a negative correlation between era total and hindsight subscore (Table~\ref{tab:summary}, Figure~\ref{fig:leak}); after removing run means, the pooled within-run correlation over 24 eras is $r=-0.58$. The direction holds for the mechanical aggregate $\tilde{T}$, even though $\tilde{T}$ contains the hindsight subscore and is biased toward a positive correlation. The per-run estimates rest on six points each; their common sign across four runs and three orchestration schemes is the robust part.

\begin{table}[t]
\centering
\small
\caption{Training-era summary (E1--E6, $n=6$ per run). Slope: least-squares change in published total per era. $h$: hindsight subscore (10 = no leakage).}
\label{tab:summary}
\begin{tabular}{lcccccccc}
\toprule
Run & Mean $T$ & SD & Slope & E1--E3 & E4--E6 & Mean $h$ & $r(T, h)$ & $r(\tilde{T}, h)$ \\
\midrule
001 & 60.0 & 6.2 & $+2.46$ & 55.0 & 65.0 & 6.00 & $-0.58$ & $-0.26$ \\
002 & 65.0 & 4.1 & $+1.43$ & 62.7 & 67.3 & 5.17 & $-0.58$ & $-0.39$ \\
003 & 62.5 & 3.6 & $+1.46$ & 60.3 & 64.7 & 5.17 & $-0.92$ & $-0.64$ \\
004 & 66.0 & 5.2 & $+2.63$ & 61.3 & 70.7 & 7.33 & $-0.38$ & $-0.38$ \\
\bottomrule
\end{tabular}
\end{table}

Two readings fit. Under \emph{learning}, the firm improves and later eras happen to be harder to judge cleanly. Under \emph{recall}, later eras sit in the model's densest knowledge of technology history, so the firm names winners more easily, scores higher and is only partly penalized. The Run 001 E6 decision shows why the readings are hard to separate: its tripwire, ``a public model at least ten times larger than Megatron that works from prompts alone'', describes GPT-3, released five months later. That decision was written by the operator, who also knew the answer. Historical eras cannot distinguish these readings; eras the model cannot remember can (Section~\ref{sec:benchmark}).

\subsection{Channels of distortion}
\label{sec:artifacts}

\paragraph{Briefing selection.} A briefing can contain only pre-era facts and still leak. Run 004's 1990 briefing foregrounds the 1989 CERN hypertext proposal as a frontier signal. The fact is admissible; choosing it is not neutral. Because the judge writes the next briefing after seeing how the era turned out (Figure~\ref{fig:loop}), this channel is built into the loop.

\paragraph{Self-judging.} In every run the judge is the same model as the players, and in Run 004 the same root context wrote the board decisions and scored them. LLM judges prefer their own outputs \citep{panickssery2024selfpref}. In Run 001 the E8 judge's kill case became the premise of the E9 world, so forecast eras partly score the judge's own scenario.

\paragraph{The rubric shapes the firm.} Counting hindsight discipline toward the total rewards saying little about the future. The caution attractor of Section~\ref{sec:caution} is partly a product of this choice.

\begin{figure}[t]
\centering
\includegraphics[width=0.46\linewidth]{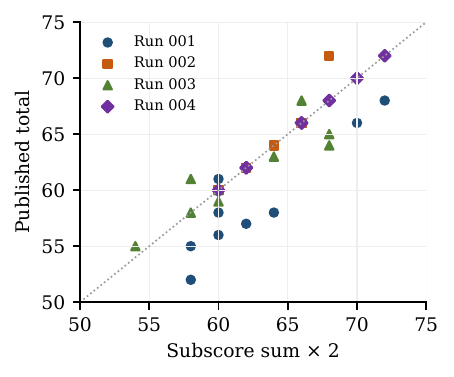}
\caption{Published total against $\tilde{T}$ for all 36 eras. Points on the dotted line follow $\tilde{T}$ exactly.}
\label{fig:agg}
\end{figure}

\paragraph{Aggregation.} The published total follows different rules in different runs (Figure~\ref{fig:agg}). Run 001 totals are holistic and lie below $\tilde{T}$ in eight of nine eras (mean $-3.7$); Run 004 uses $\tilde{T}$ exactly; Run 002 matches except E6; Run 003 matches only E1 and E2. Recomputed training means are 63.7, 64.3, 62.3 and 66.0, a narrower spread than the published 60.0--66.0. Cross-run comparisons of published totals mix rules.

\paragraph{Export errors and contamination.} In Runs 002 and 003 the live and forecast subscores were exported one column to the left; totals were unaffected, and the released data are corrected. Run 002 reuses Run 001's company names and theses for E1--E4 and restates its headline finding, and Run 004 also begins as ``Switchyard Systems'': the later runs were produced where Run 001's records were probably visible and are not independent replications.

\section{Discussion}

\paragraph{For builders of agent organizations.} Structure changes long-run behaviour in ways short tasks cannot reveal. Critics are added to agent systems to reduce errors \citep{du2024debate,liang2024divergent}; over fifty simulated years a critic with hard numeric gates made the firm unable to act. Memory is added to improve performance; its content determined what kind of firm the system became. Critics should be calibrated against the cost of inaction as well as the cost of error, and memory should be audited for what it stores, not only whether it helps.

\paragraph{For organizational theory.} LLM firms reproduced several documented regularities of human firms: perceiving distant opportunities while committing near existing competences \citep{cohen1990absorptive,march1991exploration}, drifting to timid choices under loss-averse evaluation \citep{kahneman1993timid}, and being governed by the routines their experience was encoded into \citep{levitt1988organizational}. Agent organizations could become a cheap laboratory in which such hypotheses are manipulated directly, by changing the charter, the critic or the memory, provided the leakage problem is controlled.

\paragraph{For evaluation.} Rising scores are not evidence of learning when the player and the scorer both know the answer. Any historically scored agent evaluation should report a leakage measure alongside performance, separate the briefing author from the scorer, compute totals in code, and fix the rubric's treatment of caution in advance. Each of the distortions we found was large enough on its own to change a naive comparison between runs.

\section{Toward a benchmark}
\label{sec:benchmark}

We specify the controlled version, which the released harness partly implements.

\paragraph{Splits.} \emph{Historical} eras (E1--E6) measure behaviour under a known leak. A \emph{fictional} era with an invented but consistent technology history measures the judge's false-positive leakage rate. A \emph{post-cutoff} era after the player model's training data \citep{cheng2024dated} separates foresight from recall.

\paragraph{Conditions.} Full organization, no Playbook, no Red Team, and a single-prompt founder baseline; five to ten runs each at temperature 1.0; at least two player models and a judge from a different model family, blinded to condition; every run in a clean workspace containing only prompts and harness.

\paragraph{Metrics.} $\tilde{T}$ computed by the harness from structured subscores, with hindsight reported separately and excluded; training-era slope; foresight--commitment gap; \emph{frontier capture} (whether the chosen layer matched, neighboured or missed where value pooled); pivot rate; dissent accuracy over all logged dissents; leakage from the judge and from a vocabulary probe that flags terms first used after $t_k$; and agreement with two to three human raters on a stratified sample of eras.

\paragraph{Pre-registered hypotheses.} H1: the training-era slope is higher with the Playbook than without it. H2: the full organization beats the single-prompt baseline. H3: on the fictional era, judged leakage does not differ between conditions. H4: the foresight--commitment gap is positive in all conditions. H5: removing the Red Team raises pivot rate and frontier capture.

The design needs about 46 model calls per run, or 900--1,850 calls across conditions, plus human rating time.

\section{Limitations}

The four trajectories are exploratory and use only two models; later runs were probably exposed to the first. No run instantiated sixteen independent agents. All subscores, including the two that define the foresight--commitment gap, are judge-assigned with hindsight, and qualitative labels such as ``named'' and ``adjacent'' are the judge's. Outcomes such as revenues and exits are the judge's calibrated guesses and compound across eras. Runs 002 and 003 have thinner records. Live-era audits reflect web sources as of September 2026. Forecast eras have no ground truth. Per-run correlations rest on six points.

\section{Conclusion}

We placed LLM agent organizations in fifty years of technological change and watched what they became. They saw the coming shifts and built beside them, in every historically scored era of every run. A strong critic froze one firm for half a century while its scores stayed the highest, and the content of each firm's memory set its character. Whether any of this reflects learning cannot be read from the scores, which rose as hindsight crept in. Long-horizon evaluation of agent organizations is possible and revealing, but only with eras the model cannot remember and scores it does not assign to itself. We release the testbed and a design for doing it properly.

\section*{Ethics statement}

Simulated companies, financings and outcomes are fictional and are not claims about real firms; real companies appear only in the judge's account of history or of the current market. One board decision in Run 001 was written by the human operator and is marked throughout. Three Run 001 passages that applied the simulation to one founder's personal planning were removed before release; no other text was edited.

\section*{Data and code availability}

All prompts, run records, scores (with corrected exports and a unified table), figures and the API harness are released under the MIT license \citep{frontierautolab2026repo}.

\bibliographystyle{plainnat}
\bibliography{refs}

\appendix
\section{Full score table and subscores}
\label{app:scores}

Table~\ref{tab:all} lists the published total $T$ and the mechanical aggregate $\tilde{T}$ for every era of every run. Subscores are in the repository file \texttt{results/all\_runs\_scores.csv}.

\begin{table}[H]
\centering
\small
\caption{Published total $T$ and $\tilde{T} = 2\sum_j s_j$ for all 36 eras.}
\label{tab:all}
\begin{tabular}{l*{8}{c}}
\toprule
& \multicolumn{2}{c}{Run 001} & \multicolumn{2}{c}{Run 002} & \multicolumn{2}{c}{Run 003} & \multicolumn{2}{c}{Run 004} \\
\cmidrule(lr){2-3}\cmidrule(lr){4-5}\cmidrule(lr){6-7}\cmidrule(lr){8-9}
Era & $T$ & $\tilde{T}$ & $T$ & $\tilde{T}$ & $T$ & $\tilde{T}$ & $T$ & $\tilde{T}$ \\
\midrule
E1 1990 & 55 & 58 & 62 & 62 & 60 & 60 & 60 & 60 \\
E2 1996 & 52 & 58 & 66 & 66 & 58 & 58 & 62 & 62 \\
E3 2002 & 58 & 64 & 60 & 60 & 63 & 64 & 62 & 62 \\
E4 2008 & 66 & 70 & 66 & 66 & 65 & 68 & 70 & 70 \\
E5 2014 & 68 & 72 & 64 & 64 & 61 & 58 & 70 & 70 \\
E6 2020 & 61 & 60 & 72 & 68 & 68 & 66 & 72 & 72 \\
\midrule
E7 2026 & 56 & 60 & 66 & 66 & 64 & 68 & 66 & 66 \\
E8 2032 & 57 & 62 & 62 & 62 & 59 & 60 & 68 & 68 \\
E9 2040 & 58 & 60 & 60 & 60 & 55 & 54 & 66 & 66 \\
\bottomrule
\end{tabular}
\end{table}

\begin{figure}[H]
\centering
\includegraphics[width=0.95\linewidth]{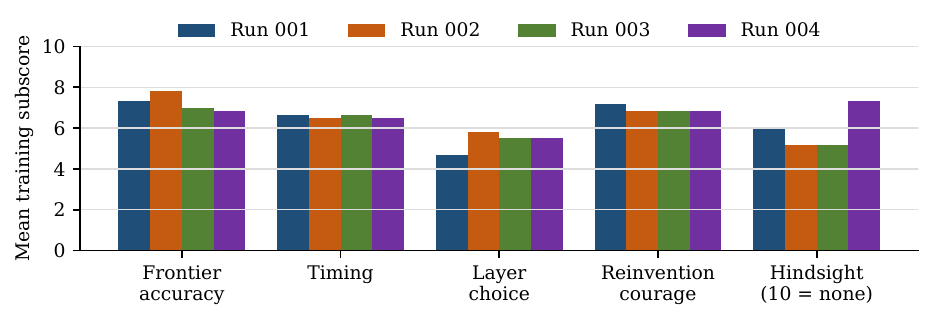}
\caption{Mean training-era subscores by run.}
\end{figure}

\section{Company identities by run}
\label{app:identity}

\begin{table}[H]
\centering
\small
\caption{Company identity chosen by the board in each era.}
\begin{tabular}{lllll}
\toprule
Era & Run 001 & Run 002 & Run 003 & Run 004 \\
\midrule
E1 1990 & Switchyard Systems & Switchyard & Porthole Networks & Switchyard Systems \\
E2 1996 & Manifest Networks & Manifest & PageSignal & Switchyard Systems \\
E3 2002 & Ledgerline & Ledgerline & Clinisphere & Switchyard Commerce Ops. \\
E4 2008 & Clearline & Clearline & ClaimGraph & Switchyard Commerce Ops. \\
E5 2014 & Clearsight & Vectorial & BenefitFlow & Switchyard Evidence Ops. \\
E6 2020 & Clearproof & Proofline & LineSight & Switchyard Evidence Ops. \\
E7 2026 & Proofworks & Tracewell & FlowCheck & Caseground \\
E8 2032 & Clearwork & Consequence & WorkPermit & Caseground \\
E9 2040 & Clearbond & Recourse & Delegation Warranty & Caseground \\
\bottomrule
\end{tabular}
\end{table}

\section{Judge instructions (excerpt, Run 001)}

The training-era judge was instructed to describe ``what actually happened in the real world during this era window (real winners, the real bottleneck or layer where value pooled, the real capability jumps)'', to score five dimensions from 0 to 10 with one-line justifications, including ``hindsight leakage (10 = no leakage; penalize any post-date knowledge in memos or decision)'', to give an overall era score from 0 to 100 with a calibrated simulated outcome, and to append three to five generalizable lessons to the Playbook. It then wrote the next era's briefing, restricted to information public before that era's start date. Full prompts are in the repository under \texttt{prompts/}.

\end{document}